\documentclass[pra,10pt,aps,twocolumn,amsmath,amssymb,superscriptaddress,altaffilletter]{revtex4-2}

\usepackage{lmodern}
\usepackage{graphicx}
\usepackage{dcolumn}
\usepackage{bm}
\usepackage[utf8]{inputenc}
\usepackage[T1]{fontenc}
\usepackage{color}
\usepackage[colorlinks,citecolor=blue,linkcolor=blue,urlcolor=blue,bookmarks=false,hypertexnames=true]{hyperref}

\begin{document}

\title{Attraction Induced by Mutual Quantum Measurements of Velocity}

\author{Walter Hahn}
\email{walter.hahn@uibk.ac.at}
\affiliation{Skolkovo Institute of Science and Technology, Skolkovo Innovation Centre, Nobel Street 3, Moscow 143026, Russia}
\affiliation{Institute for Quantum Optics and Quantum Information of the Austrian Academy of Sciences,  Innsbruck,  Austria}

\author{Boris V. Fine}
\email{fine.bv@mipt.ru}
\affiliation{Skolkovo Institute of Science and Technology, Skolkovo Innovation Centre, Nobel Street 3, Moscow 143026, Russia}
\affiliation{Laboratory for the Physics of Complex Quantum Systems, Moscow Institute of Physics and Technology,Institutskiy per. 9, Dolgoprudny, Moscow region, 142290, Russia}
\affiliation{Institute for Theoretical Physics, University of Heidelberg, Philosophenweg 12, 69120 Heidelberg, Germany}


\begin{abstract}
We define the notion of mutual quantum measurements of two macroscopic objects and investigate the effect of these measurements on the velocities of the objects. We show that multiple mutual quantum measurements can lead to an effective force emerging as a consequence of asymmetric diffusion in the velocity space. We further show that, under a certain set of assumptions involving the measurements of mutual Doppler shifts, the above force can reproduce Newton's law of gravitation. For a broader class of measurements, the emergent force can also lead to corrections to Newton's gravitation. 
\end{abstract}

\maketitle

\section{Introduction} \label{sec_intro}

We investigate the effects of mutual quantum measurements of two macroscopic objects. The quantum measurements are assumed to proceed through an entanglement and decoherence process mediated by the exchange of particles moving with the speed of light. The measurements register the Doppler shifts of the exchanged particles. We show that such a process can lead to an attraction originated from the velocity diffusion caused by the measurements. Such an attraction might be the mechanism behind Newton's gravitation, or it might cause corrections to Newton's gravitation.

Some of the ideas presented in this article about the interplay of quantum mechanics and gravitation resonate with those described in Refs.~\cite{verlinde,stamp,giulini,automaton,pikovski,lloyd,Diosi-90,Penrose1996,kent2,Altamirano_2017,brukner,grwp,diosi,Kafri_2014,Kafri_2015,nimmrichter,tilloy,khosla2018classical}. However, in these references, the authors adopt the perspective of either high-energy physics, or black-hole physics, or consider gravity as given and look for the consequences of it in terms of quantum properties. Here, in contrast, we present a consistently low-energy non-relativistic perspective, which puts emphasis on routine quantum measurements as a possible mechanism of gravitation. The speed of light only appears in the first-order Doppler frequency shift, which has the same form in both relativistic and non-relativistic theories. Our plan is to apply non-relativistic quantum mechanics as far as possible towards the Planck spatial scale subject only to quantum measurements. A more detailed comparison of the present work with some of the other proposed treatments is given in Sec.~\ref{discussion}.

\section{Preliminary discussion} \label{preliminary}

If a macroscopic object is completely isolated from the environment, its internal degrees of freedom are decoupled from its center of mass (CM) and hence cannot induce quantum decoherence of the CM wave function. The CM wave function loses coherence as a result of the interaction with the environment~\cite{joos_zeh,Joos1985,zurek,darwinism}, e.g., through emission or absorption of a photon~\cite{arndt}.

The decoherence of the CM wave function can be induced by entangling external particles with CM positions or CM momenta, or both. As a result, the CM density matrix can be thought of as a mixture of coherent wave packets, each having a certain stationary coherence width in both real space and momentum space~\cite{giulini,joos_zeh}. Both widths are likely rather small, but once they are reached, the decoherence process is supposed to be compensated by a dynamic expansion of the wave packets. 

Thermal de Broglie wave length can serve as  initial estimate of the above coherence width in real space. Such an estimate indicates that, for any reasonable temperature, the CM wave functions of macroscopic objects are localized on subnuclear scales. This makes the entanglement of surrounding microscopic particles with the CM position rather inefficient. Likewise, the stationary coherence width in the momentum space might be large in absolute terms, but, once divided by mass, gives velocity distributions which are very narrow and hence are difficult to discriminate through emission or scattering of microscopic particles.

In this article, we focus on the processes that, despite the above inefficiency, limit the coherence width of the CM wave packets in the momentum space. We explore a scenario, in which the leading process  controlling the coherent momentum width of a macroscopic object is the small Doppler shift of the emitted photons. This shift leads to orthogonal photon states that become entangled with another macroscopic object some distance away, and hence the velocity of the former object becomes, in effect, measured. 

Below, for the sake of conceptual simplicity, we describe some of the processes in the language of wave functions, but it is possible to reformulate our description in the language of density matrices. We also often use a crude coarse-graining of the wave packets considered. The resulting description, however, should be amenable to mathematical derivations in continuous space and time. 

\section{Mutual Quantum measurements} \label{mutual}

\subsection{System} \label{system}

We consider two macroscopic objects  of masses $M_\text{A}$ and $M_\text{B}$ that can emit and absorb photons and also interact with an environment of microscopic particles around them. The surrounding particles have negligible masses  and energies but can cause quantum decoherence. We do not include the gravitational interaction between particles at the level of the formulation of the problem, but rather investigate whether it can emerge as a result of mutual measurements.

\subsection{Single measurement} \label{single}

In this part, we describe a single event of object B measuring the momentum of object A. 
Object A is to be represented by the wave function $ |f\rangle \otimes |F\rangle$, and object B by $ |s\rangle \otimes |S\rangle$, where $ |f\rangle$ and $ |s\rangle$ represent the respective CM wave packets in the momentum space, while $ |F\rangle$ and $ |S\rangle$ are many-body wave functions representing the internal degrees of freedom of the respective objects. 

\begin{figure}[t]
 \centering
 \includegraphics[width=0.9\columnwidth]{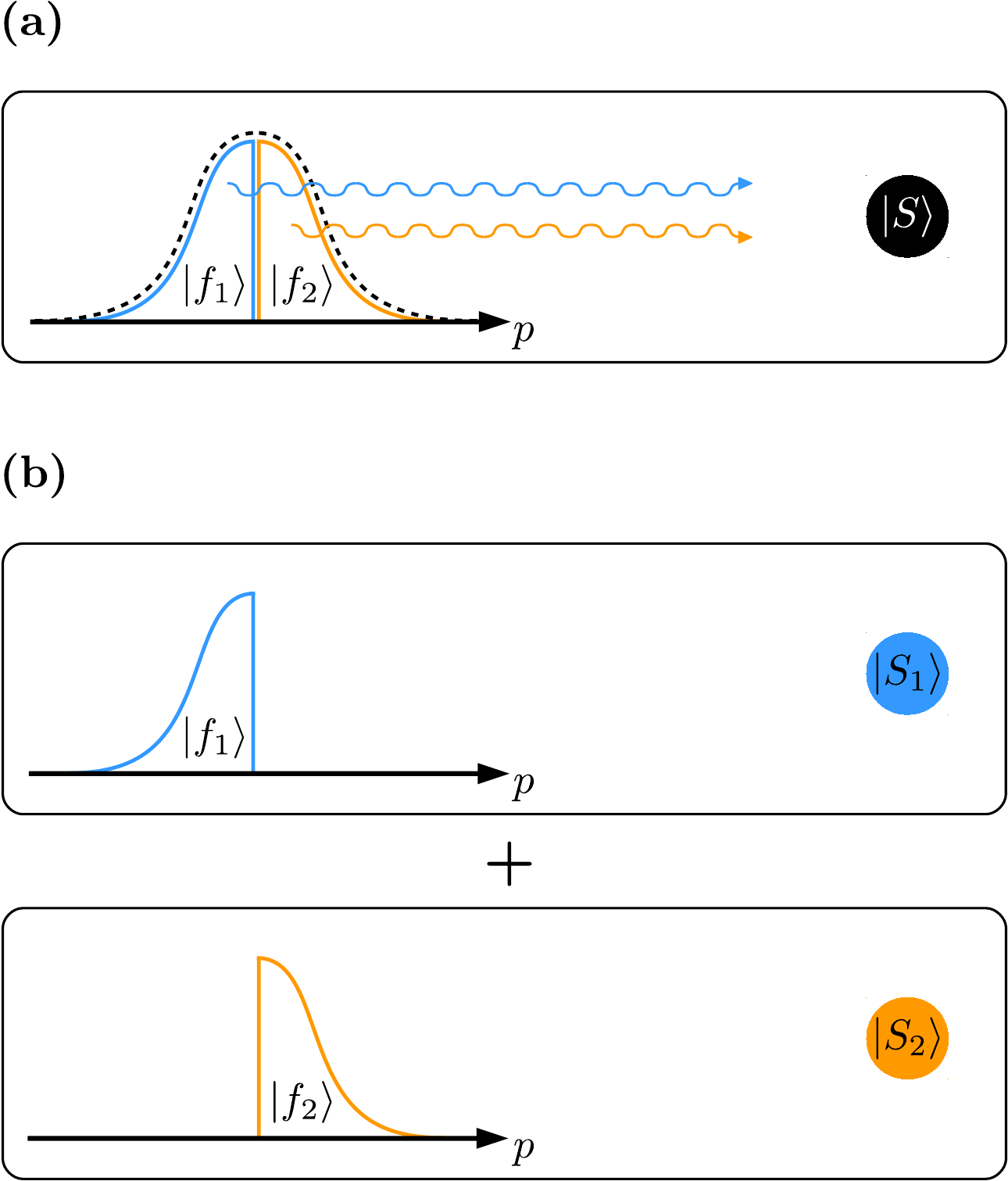}
 \caption{Illustration of the entanglement process between the center of mass of object A (states $|f_1\rangle$ and $|f_2\rangle$) and the internal degrees of freedom of object B (states $|S\rangle$, $|S_1\rangle$ and $|S_2\rangle$): \textbf{(a)} Beginning of the process, as represented by Eq.~(\ref{Psi2}); \textbf{(b)} Final entangled state given by Eq.~(\ref{Psi3}).}
 \label{fig_psi}
\end{figure}

Let us consider a process illustrated in Fig.~\ref{fig_psi}, by means of which the CM momentum of object A becomes entangled with the microscopic degrees of freedom of object B. We represent the initial momentum wave packet of object A as a superposition of its two parts $|f_1\rangle$ and $|f_2\rangle$ shown in Fig.~\ref{fig_psi}(a), so that the initial state of the total system is
\begin{equation}
|\Psi_1\rangle = \frac{1}{\sqrt{2}} \Big(|f_1\rangle + |f_2\rangle\Big) \otimes |F\rangle \otimes |s\rangle \otimes |S\rangle   .
\label{Psi1}
\end{equation}
Object A then emits a photon towards object B. Here and below we assume that the recoil due the emission or absorption of a photon can be neglected --- subject to the consistency check in Section~\ref{discussion}. The photon, however, is entangled with the CM momentum of object A via a Doppler-shift mechanism to be discussed later.  As a result of this emission, the internal state of object A changes to $|F'\rangle$, while the wave function of the total system becomes
\begin{equation}
|\Psi_2\rangle = \frac{1}{\sqrt{2}} \Big(|f_1\rangle  \otimes |P_1 \rangle  + |f_2\rangle  \otimes |P_2 \rangle\Big) 
\otimes |F' \rangle \otimes |s\rangle \otimes |S\rangle   ,
\label{Psi2}
\end{equation}
where $|P_1 \rangle$ and $|P_2 \rangle$ are orthogonal photon states, see Fig.~\ref{fig_psi}(a). As further illustrated in Fig.~\ref{fig_psi}(b), the photon is later absorbed by object B changing its internal state, respectively, to either $|S_1 \rangle$ or $|S_2 \rangle$, which gives
\begin{equation} \label{Psi3}
|\Psi_3\rangle = \frac{1}{\sqrt{2}} \Big(|f_1\rangle  \otimes |S_1 \rangle  + |f_2\rangle  \otimes |S_2 \rangle\Big)\otimes |F' \rangle \otimes |s\rangle.
\end{equation}
Since $|f_1\rangle$ and $|f_2\rangle$ are orthogonal to each other, $|S_1 \rangle$ and $|S_2 \rangle$ are also mutually orthogonal. As a result, the CM momentum of object A becomes entangled with the internal state of object B, which plays the role of a macroscopic observer.

We assume that the wave functions $|S_1 \rangle$ and $|S_2 \rangle$ quickly evolve to become macroscopically distinct (Schr{\"o}dinger-cat-like~\cite{Schroedinger}), so that the superposition appearing in Eq.~\eqref{Psi3} loses coherence on the time scale, which we consider infinitely short (see Appendix~A for further discussion). As a result, the density matrix describing combined properties of the CM of object A and the internal degrees of freedom of object B in the basis of quantum states $\{ |f_1 S\rangle, |f_2 S\rangle, |f_1 S_1\rangle,  |f_2 S_2\rangle\}$  evolves  from the initial one corresponding to the wave function~\eqref{Psi1} and representing an uncorrelated pure quantum state
\begin{equation} \label{rho0}
\rho_\text{init}=
\begin{array}{ l | c c c c}
&   \langle f_1 S | &  \langle f_2 S | & \langle f_1 S_1 | & \langle f_2 S_2 | \\
\hline
|f_1 S\rangle &  1/2  & 1/2  & 0 & 0 \\
|f_2 S\rangle &  1/2  & 1/2  & 0 & 0 \\
|f_1 S_1\rangle &  0  & 0  & 0 & 0 \\
|f_2 S_2\rangle &  0  & 0  & 0 & 0
\end{array}
\end{equation}
to the final one representing a classically correlated mixed state
\begin{equation} \label{rhof}
\rho_\text{final}=
\begin{array}{ l | c c c c}
&   \langle f_1 S | &  \langle f_2 S | & \langle f_1 S_1 | & \langle f_2 S_2 | \\
\hline
|f_1 S\rangle &  0  & 0  & 0 & 0 \\
|f_2 S\rangle &  0  & 0  & 0 & 0 \\
|f_1 S_1\rangle &  0  & 0  & 1/2 & 0 \\
|f_2 S_2\rangle &  0  & 0  & 0 & 1/2
\end{array}.
\end{equation}

Once the density matrix~\eqref{rhof} becomes diagonal in the chosen basis, its diagonal elements can be considered as classical probabilities and the subsequent evolution can be obtained by the appropriate random sampling of either state $ |f_1 S_1\rangle$ or state $|f_2 S_2\rangle$. One can view this sampling either as a mathematical tool or as a reflection of the actual collapse of the quantum state to a particular realization of reality. The distinction between the two interpretations is a delicate issue discussed in Appendix~A.

Below we adopt the language of collapse as more intuitive for the subsequent discussion~\cite{neumann}. We say that, once the density matrix~\eqref{rhof} becomes diagonal, the observer $| S \rangle$ has \textit{measured} the CM state $| f \rangle$ and, as a result, one of the two states $ |f_1 S_1\rangle$ or $|f_2 S_2\rangle  $ emerges. Technically, the description of the system is then continued by randomly choosing either the state $|f_1 S_1\rangle$ or $|f_2 S_2\rangle$ (postselection). For simplicity, we limited the above discussion to strong measurements, but it is possible to consider weak measurement in the more general framework of positive-operator-valued measures (POVM)~\cite{nielsenchuang}.

Eventually, the momentum of object A becomes measured by that object's internal degrees of freedom. Such a measurement can be realized once object B emits towards object A a photon that would discriminate states $| S_1 \rangle$ from $| S_2 \rangle$ and, hence,  $| f_1 \rangle$ from $| f_2 \rangle$.

\subsection{Multiple measurements vs. wave packet expansion} \label{multiple}

If the position or momentum of an object is measured repetitively, then, in the absence of a wave-packet expansion, multiple measurements would gradually reduce the width of the wave packet. This width collapse is, however, supposed to be eventually counteracted by a wave packet expansion. In particular, the position wave packets of free particles in real space just expand ballistically on their own (see Appendix~\ref{sec_spreading}). On the contrary, the momentum wave packets of free particles do not expand with time, which is the consequence of momentum conservation. They, however, can expand due to interaction with the external world that can be broadly divided into two categories (i) position- or momentum-dependent force acting on the wave packet and (ii) environment-induced measurements of the CM position, which increase the momentum width because of the Heisenberg uncertainty relation. 

In the following, we assume that we are dealing with the stationary situation, where the expansion of momentum wave packets of both massive objects occurring between the emissions of photons is compensated by narrowing caused by momentum measurements, cf. Ref.~\cite{hahn_measure}.

\section{Multiple velocity measurements and emergent attraction} \label{multiple_velocity}

We now switch the focus from momenta to velocities and concentrate on the effects of multiple velocity measurements. We denote the CM velocities of the objects A and B as $V_f \equiv p_f/M_\text{A}$ and $V_s \equiv p_s/M_\text{B}$ in terms of their respective momenta $p_f$ and $p_s$. We also denote the average velocities of the two CM wave packets as $V_F \equiv \langle V_f \rangle $ and $V_S \equiv \langle V_s \rangle$. 

\subsection{Object B measuring object A} \label{second}

In this subsection, we call object A a ``source'' and object B an ``observer''.

In order to describe the evolution of the velocity distribution of the  source as measured by the observer, we adopt the following model: The source emits towards the observer photon wave packets carrying information about the velocity of the source with root-mean-squared (rms) resolution $\Delta v_\text{A}$. The observer receives one unit of this information per time $\tau$. Our assumption of the stationary coherence width of the wave packet implies that, if no velocity measurement is made during time interval $\tau$, the mean-squared uncertainty of observer's information about the source velocity would increase to $2 \Delta v_\text{A}^2$ due to a wave packet expansion mechanism to be discussed later.

Let us denote the velocity of the CM of the source relative to the observer as $v \equiv V_f - V_S$. The observer is supposed to register a random walk of the measured value of  $v$, which then should lead to the \textit{velocity diffusion} described by the following equation for the velocity probability distribution $P(t,v)$:
\begin{equation} \label{dPdt}
\frac{\partial P}{\partial t } = \frac{\partial}{\partial v}\left(D_{\text{v}} \frac{\partial P}{\partial v}\right),
\end{equation} 
where  $D_{\text{v}} = \frac{1}{2} \Delta v_\text{A}^2/\tau$ is the velocity diffusion coefficient.

We now observe an important fact, namely, that, if the source moves with velocity $v$ towards the observer and emits photons with constant rate $1/\tau_\text{A}$ in its rest frame, then  the observer receives the photons with a different rate, namely, the one modified by the Doppler effect. In the  first order in the ratio $v/c$, the rate at which the observer becomes entangled with the velocity of the source is \ $\dfrac{1}{\tau} = \dfrac{1}{\tau_\text{A}} \left(1 - \dfrac{v}{c}\right)$. Here and below, the positive direction of velocities is from object B to object A. 

As a result of the Doppler shift, the velocity diffusion coefficient $D_{\text{v}}$ acquires the velocity dependence
\begin{equation}
D_{\text{v}}(v) = \frac{1}{2} \frac{\Delta v_\text{A}^2}{\tau_\text{A}} \left(1-\frac{v}{c}\right),
\label{Dv-shifted}
\end{equation}
which leads to the drift of the average velocity to the region, where $D_{\text{v}}$ is larger. Such a drift, in turn, implies that the observer registers a non-zero average acceleration of the source 
\begin{equation}
a_\text{A} \equiv \frac{d \langle v \rangle}{dt}.
\label{a1def}
\end{equation}

The drift of the average velocity $\frac{d \langle v \rangle}{dt}$ can be obtained by integrating the probability flux $- D_{\text{v}} \frac{d P}{dv}$ along the velocity axis
\begin{eqnarray}
\frac{d \langle v \rangle}{dt}  &=&  - \int^{+\infty}_{-\infty} D_{\text{v}} \frac{d P(v)}{d v}dv \nonumber \\
&=&  \int^{+\infty}_{-\infty}\frac{d D_{\text{v}}}{dv} P(v)dv  \approx  \left. \frac{d D_{\text{v}}}{dv} \right|_{v = \langle v \rangle}, 
\label{dvdt-integral}
\end{eqnarray}
where in the last step we assumed that $\frac{d D_{\text{v}}}{dv}$ varies very slowly across the width $w$ of the probability distribution $P(v)$ at a given time $t$. The width $w$ is defined through the velocity variance as $w^2 \equiv \langle (v - \langle v \rangle)^2 \rangle$. The condition on the slow variation of $D_{\text{v}}$ is $w \, \frac{d^2 D_{\text{v}}}{dv^2}/\frac{d D_{\text{v}}}{dv} \ll 1$.   This condition is satisfied, because, according to \eqref{Dv-shifted}, 
\mbox{$\frac{d^2 D_{\text{v}}}{dv^2} = 0 $}. In turn, the approximation \eqref{Dv-shifted} requires that
$\langle v \rangle \ll c$ and $w \ll c$.

Substituting Eq.~\eqref{Dv-shifted} into Eq.~\eqref{dvdt-integral} and then using the definition~\eqref{a1def}, we obtain
\begin{equation} \label{a1}
a_\text{A} = - \frac{1}{2 c} \frac{\Delta v_\text{A}^2}{\tau_\text{A}}.
\end{equation}
The negative sign in the right-hand side of Eq.~\eqref{a1} implies that the acceleration is in the direction from object A to object B, which corresponds to attraction.

It is important to mention that, in addition to the drift of the average velocity $\langle v \rangle$, the velocity diffusion described by Eq.~\eqref{dPdt} also leads to the growth of the velocity variance $w^2$.  From Eqs.~(\ref{dPdt},\ref{Dv-shifted}), we obtain, in the leading order in $v/c$,
\begin{equation}
 \frac{d w^2(t)}{dt}= 2 \langle D_\text{v}\rangle.
 \label{dw2}
\end{equation}
Since the growth of the velocity variance implies the growth of the average kinetic energy, we refer to this process as ``heating". In principle, the growth of $w$ can be significant in comparison with the growth of $\langle v \rangle$ , because, according to Eq.~(\ref{Dv-shifted}), the former  is of the order of 1, while the latter is of the order $v/c$\footnote{We thank Lajos Diosi for pointing out this important issue.}. However, in the limit $w \ll c$, the growth of $w$ does not affect the evolution of $\langle v \rangle (t)$.  The heating issue is to be addressed further in Section~\ref{heating}. 

\subsection{Object A measuring object B and the combined effect} \label{first}
The preceding derivation can be repeated for the apparent acceleration $a_\text{B}$ of the center of mass of object B ``observed'' by the internal degrees of freedom of object A. Denoting the rms velocity fluctuations of the CM of object B as $\Delta v_\text{B}$ and the characteristic photon emission time as $\tau_\text{B}$,  we obtain 
\begin{equation}
a_\text{B} = \frac{1}{2c}\frac{\Delta v_\text{B}^2}{\tau_\text{B}}.
\label{a2}
\end{equation}
The positive sign $a_\text{B}$ implies the acceleration from object B to object A.

The considerations leading to Eqs.~\eqref{a1} and~\eqref{a2} can now be summarized as follows: The internal degrees of freedom of object B ``register'' the diffusion of the average CM velocity $V_F$ of object A, while the internal degrees of freedom of object A register the diffusion of the average CM velocity $V_S$ of object B. Thus, the combined internal degrees of freedom of the two objects register that both ``ends'' of the expression for the relative average velocity, $V_F - V_S$, undergo independent diffusion with a systematic drift towards each other. As a result, the total relative acceleration of the two objects as recorded by their internal degrees of freedom is 
\begin{equation}
a_\text{A} - a_\text{B} = - \frac{1}{2 c} \left( \frac{\Delta v_\text{A}^2}{\tau_\text{A}} +\frac{\Delta v_\text{B}^2}{\tau_\text{B}} \right).
\label{a1a2}
\end{equation}

\section{Assumptions leading to Newton's gravitation} \label{assumptions}

Here we propose a set of ideas about possible mechanisms controlling the right-hand-side of Eq.~\eqref{a1a2} and make estimates based on these mechanisms. 

\subsection{Typical amplitude of velocity fluctuations $\Delta v_\text{A}$} \label{TypicalAmplitude}

Object B measures the velocities within the coherent velocity wave packet of object A by becoming entangled with that wave packet. We assume that the main mechanism of this entanglement is the detection of Doppler shifts of the photons emitted from different parts of object's A velocity wave packet [for example, states $|f_1 \rangle$ and $|f_2 \rangle$ in Eq.~(\ref{Psi1})]. Below, we estimate the velocity resolution $\Delta v_\text{A}$ due to such a mechanism.

\begin{figure}[t]
 \centering
 \includegraphics[width=0.99\columnwidth]{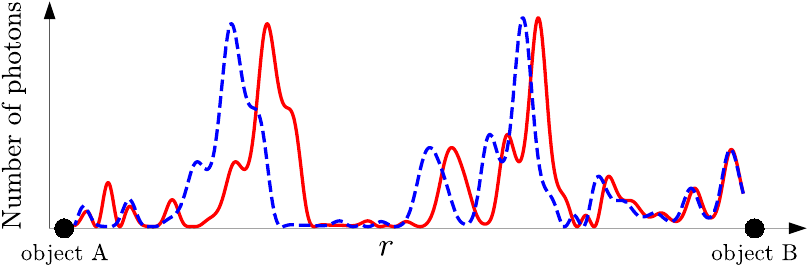}\llap{\makebox[7.5cm][l]{\raisebox{2cm}{\includegraphics[height=0.5cm]{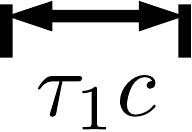}}}}
 \caption{Train of photonic wave packets emitted by object A and moving from object A to object B. This sketch illustrates the derivation of condition~(\ref{resolve}). The blue dashed line represents the configuration where object A is at rest, and the red solid line the configuration where object A moves with very small velocity $\Delta v_\text{A}$ towards object B.}
 \label{fig_two_photons}
\end{figure}

Our estimate is to be done for the setting where object A sends towards object B a train of non-monochromatic wave packets with characteristic correlation time $\tau_\text{A}$ (see Fig.~\ref{fig_two_photons}). This characteristic time is then converted into the characteristic length $c \tau_\text{A}$ of the wave packets. The above train of photon wave packets can propagate either on its own or as a modulation of radiation with frequency much higher than $1/\tau_\text{A}$, e.g., thermal radiation. 

Let us now consider a quantum superposition of two configurations: (i) object A is at rest with respect to object B, and (ii) object A moves towards object B with very small velocity 
$\Delta v_\text{A}$. The initial position of the CM of object A and the 
state of its internal degrees of freedom are the same in the both 
settings.  Once emitted, the wave packets arrive to object B
after time $t_0 = r/c$, where $r$ is the distance between the two 
objects. We assume $r \gg c\tau_\text{A}$.

Let us further suppose that a photon wave packet of size $c\tau_\text{A}$ is 
emitted in each of the superposed configurations at time $t=0$ in object's B rest frame. This wave packet arrives to object B at the same time $t=t_0$ for the both configurations, which means 
that object B cannot yet discriminate between the two 
configurations and hence resolve $\Delta v_\text{A}$. However, this first 
detection initializes the subsequent measurement, which then occurs as 
follows:  By the time $t=t_0$, object A emits on average 
$t_0/\tau_\text{A}$ wavepackets, when it is at rest. When it moves with 
velocity $\Delta v_\text{A}$, then, due to the Doppler shift, it emits photons 
slightly faster, namely, with rate
\begin{equation}
 \dfrac{1}{\tau'} = \dfrac{1}{\tau_\text{A}} \left(1 + \dfrac{\Delta v_\text{A}}{c}\right),
\end{equation}
which means $t_0/\tau_\text{A} (1 + \Delta v_\text{A}/c)$ emitted wave packets by the time $t=t_0$.
Object B should be able to discriminate between the two 
photonic trains when their maxima become shifted with respect to each 
other by the characteristic wave packet length $c \tau_\text{A}$,  as sketched 
in Fig.~\ref{fig_two_photons}. Further assuming that each photon wave 
packet of size $c \tau_\text{A}$ is detected with probability~1, object B should be able to resolve the velocity difference $\Delta v_\text{A}$ by 
time $t=2t_0$, if
\begin{equation}
\Delta v_\text{A} \gtrsim  \frac{ \tau_\text{A} c}{t_0 } .
\label{resolve}
\end{equation}

If the above inequality is not satisfied, and the velocity of object A does not fluctuate in time, then object B can just wait 
longer and, eventually, resolve $\Delta v_\text{A}$. However, according to the assumptions of subsection~\ref{second}, the velocity of object A 
does fluctuate due to external factors, which, in turn, leads to the 
broadening of the velocity wave packet (see also 
subsection~\ref{CharacteristicTime} below).  One can think of these 
fluctuations as yet another kind of random walk in the velocity space.   
We expect that out of, possibly, a much broader spectrum of the 
electromagnetic radiation emitted by object A, the channel that 
most efficiently communicates the value of the fluctuating velocity is 
the one carried by the stochastic component with the photon correlation time 
that matches the time scale of the velocity fluctuations.  We, 
therefore, assume that the velocity of object A fluctuates with the photon 
correlation time $\tau_\text{A}$. This time scale is to characterize both the 
fluctuations of the center of the velocity wave packet and the broadening 
of that wave packet.

Let us now follow the line of reasoning that led to condition~\eqref{resolve} but 
do it in the presence of velocity fluctuations of object A and 
for a continuous initial velocity wave packet of width $\Delta v_\text{A}$. In 
this case, the initialization stage is the same as before.  Moreover, 
the condition~\eqref{resolve} still allows object B to resolve 
the difference $\Delta v_\text{A}$ between two identical random walks of object's A velocity $v_1$ --- one starting from the initial velocity 
$v_1 =0$ and the other one from $v_1 = \Delta v_\text{A}$. However, if $\Delta 
v_1$ is smaller than the right-hand-side of inequality~\eqref{resolve} and, hence, not resolved by time $t_0$ after the initialization moment, then the 
resolution $\Delta v_\text{A}$ will never be achieved, because waiting longer will not help. The reason is that the growing velocity uncertainty will exceed the information gained by the measurement.
 The balance between the 
velocity uncertainty generated by fluctuations and the information 
aquired by object B gives the condition
\begin{equation}
\Delta v_\text{A} \simeq  \frac{ \tau_\text{A} c}{t_0}  = \frac{\tau_\text{A} c^2}{r} .
\label{resolve2}
\end{equation}
It means that the greater the distance between the two objects, the better the velocity resolution.

Finally, we note that, if, in the above picture, each of the wave packets communicating the Doppler shift contains $N$ coherent photons instead of one, then measuring one photon amounts to measuring all, which, in turn, implies that detecting more than one photon from the same random Doppler-shifted mode does not add more precision in determining $v_1$.

\subsection{Characteristic time of velocity fluctuations $\tau_\text{A}$} \label{CharacteristicTime}

Let us assume that the free motion of the CM of object A is 
modified on a certain small length scale $l_0$. The reasons can be both 
routine, such as fluctuations of the net force exerted on the object by 
the environment, or fundamental, such as a departure from the 
Schr{\"o}dinger dynamics on the Planck and sub-Planck spatial 
\mbox{scales~\cite{minimum_length,garay,adler}}.

Let us now consider a Gaussian wave packet 
\mbox{$\psi(x_\text{A})\cong\exp[-x_\text{A}^2/(2l_0^2)]$} for the CM 
position $x_\text{A}$ of object A.
If there were no correction to the Schr{\"o}dinger dynamics,  this wave 
packet would be expanding, and, as a result,
the variance of $x_\text{A}$ would double after time $\tau_\text{AG} = M_\text{A} l_0^2 
/\hbar$ (see Appendix~\ref{sec_spreading}).  But the 
momentum distribution associated with this wave packet  will continue having the same width. However, if there 
is a large correction to the Schr{\"o}dinger dynamics on the length 
scale $l_0$, then the momentum distribution characterizing the wave 
packet is expected to change significantly over time $\tau_\text{AG}$. This 
implies that the characteristic time of velocity fluctuations $\tau_\text{A}$ 
is of the order of $\tau_\text{AG}$, i.e.,
\begin{equation} \label{tau1}
\tau_\text{A} \sim \frac{M_\text{A} l_0^2}{\hbar}.
\end{equation}

The assumptions adopted in the preceding subsection imply the chaotic 
character of the dynamics underlying the velocity fluctuations. Indeed, 
if object B separated from object A by distance $r_1$ 
measures at $t=0$ the velocity of object A with uncertainty 
$\Delta v_\text{A}$ given by Eq.~(\ref{resolve2}) and, later, does not receive 
more information, then, according to our picture, by the time $t = 
\tau_\text{A}$, the velocity uncertainty for that object B will increase 
to $\sqrt{2} \Delta v_\text{A}$. However, if object B is
located at distance $r'_1 = 2 r_1$, then, according to Eq.~(\ref{resolve2}), it measures the velocity at $t=0$ with 
accuracy $\Delta v'_\text{A} = \Delta v_\text{A}/2$, and the uncertainty accumulated 
by time $t = \tau_\text{A}$ in the absence of further measurement will be 
$\sqrt{2} \Delta v'_\text{A}$. In other words, the velocity uncertainty due to 
the dynamics of object A increases by the same factor over the 
same time interval irrespective of uncertainty's initial value. Such a
property is exhibited by classical chaotic dynamics characterized by the 
Lyapunov exponent $\lambda_\text{A} \sim 1/\tau_\text{A}$.

\subsection{Planck length as the characteristic length $l_0$} \label{PlanckLength}

We now assume that the characteristic length $l_0$ for violating the free-particle Schr{\"o}dinger dynamics and, thus, causing the velocity fluctuations is independent of particle's mass, and the value of $l_0$ is of the order of the Planck length
\begin{equation} \label{lp}
l_0 \cong \sqrt{\frac{G \ \hbar}{c^3}},
\end{equation}
see Appendix~\ref{Planck} for more discussion.

Making chain substitutions of Eqs.~(\ref{resolve2}), (\ref{tau1}) and (\ref{lp}) into Eq.~(\ref{a1}), we obtain
\begin{equation} \label{a1G}
a_\text{A} \cong -  G \ \frac{M_\text{A}}{r^2},
\end{equation}
where we omit the numerical prefactor, which is uncontrollable in the present treatment. The same chain of estimates for object B gives $a_\text{B} \cong   G \frac{M_\text{B}}{r^2}$. Thus, according to Eq.~(\ref{a1a2}), the relative acceleration reads
\begin{equation}
a_\text{A} - a_\text{B} \cong -  G \ \frac{M_\text{A} + M_\text{B}}{r^2}.
\label{a1a2G}
\end{equation}
The above expression is consistent with Newton's law of gravitation, subject to the following clarification:  $a_\text{A}$ and $a_\text{B}$ are  not the  accelerations of the respective objects in the inertial laboratory reference frame, rather they are \textit{measured} relative accelerations as defined in subsection~\ref{second}.  

We now denote the resulting accelerations of the objects A and B in an inertial laboratory reference frame as $a'_\text{A}$ and $a'_\text{B}$ and postulate that $a'_\text{A} - a'_\text{B} = a_\text{A} - a_\text{B}$. We further require the total momentum conservation in the form $M_\text{A} a'_\text{A} = -M_\text{B} a'_\text{B}$. These two postulates together with Eq.~(\ref{a1a2G}) lead to an expression consistent with the standard form of Newton's law of gravitation:
\begin{equation} \label{a1a2prime}
M_\text{A} a'_\text{A} = -M_\text{B} a'_\text{B} \cong - G \ \frac{M_\text{A} M_\text{B}}{r^2}.
\end{equation}
Let us now observe that, in the preceding derivation, masses $M_\text{A}$ and $M_\text{B}$ were defined as inertial masses. Their appearance in Eq.~\eqref{a1a2prime} as gravitational masses originates from the assumption that the time scale of the relevant velocity fluctuations can be estimated by Eq.~(\ref{tau1}) as the characteristic time of the inertial broadening of a spatial wave packet of a free massive quantum particle. Thus, the present treatment guarantees the equivalence between gravitational and inertial masses.

\section{Measurement-induced heating}
\label{heating}

The main preoccupation of the present article is the measurement-induced drift of the average velocity of a wave packet. However, as mentioned at the end of subsection~\ref{second}, the proposed measurement scheme would also induce the growth of the velocity dispersion and hence the growth of the kinetic energy of the measured object, which we call ``heating''. If our model is to be considered as a mechanism behind Newtonian gravity, then the measurement-induced velocity drift $a \Delta t$ can easily become comparable to a typical velocity fluctuation $w (\Delta t) $. Since no such significant velocity fluctuations are observed to accompany the Newtonian gravity, our model would need to be supplemented by additional ingredients that lead to the suppression of heating. Below we sketch one such a possible addition.

Let us start by reiterating the observation made in subsection~\ref{second} that the growth of the velocity variance $w^2$ does not interfere with the velocity drift $\langle v \rangle$ as long as $w \ll c$. 
Let us further note that the velocity fluctuations observed by a distant observer occur on the top of the dynamics induced by the environment of the measured object. Now, let us imagine that the measured object is a large Brownian particle moving in a fluctuating environment of other Brownian particles.

The probability distribution of the measured velocity fluctuations of the above Brownian particle can described by a Fokker-Planck equation
\begin{equation}
    \frac{\partial P}{\partial t } = \frac{\partial}{\partial v}\left(\gamma[v-v_0] P +D_{\text{v}} \frac{\partial P}{\partial v}\right),
    \label{FP-gamma}
\end{equation}
where, $D_{\text{v}}$ is the velocity diffusion coefficient given by Eq.~(\ref{Dv-shifted}), $\gamma$ is the friction coefficient and $v_0$ is the velocity of the environment. The friction term limits the growth of the velocity variance to 
\begin{equation}
 w_0^2 \equiv \langle (v- v_0)^2 \rangle = \langle D_{\text{v}}\rangle / \gamma.
\label{w0}
\end{equation}
If the environment has fixed velocity $v_0$, e.g. $v_0=0$, then  it would suppress not only the velocity fluctuations but also the measurement-induced velocity drift, i.e. the acceleration. Instead of the computed acceleration $a_\text{A}$ towards the observer, the observed particle would move with a stationary velocity $a_\text{A}/ \gamma$. 

Now comes the last step: Since the observer is involved in the mutual measurement process not only with the original Brownian particle but also with the surrounding particles, then, as described in Section~\ref{PlanckLength}, all the environmental particles are supposed to experience the same acceleration towards the observer as the original one --- consequence of the equivalence principle. As a result, the environment velocity $v_0$ in Eq.~(\ref{FP-gamma}) acquires the same time-dependence as the velocity drift $\langle v \rangle$ of the selected Brownian particle.

Bringing all above elements together, the issue of heating can be resolved, if Eq.~(\ref{dPdt}) is replaced with Eq.~(\ref{FP-gamma}) with  $v_0 (t) = \langle v \rangle (t)  = \int v P(v) \, dv $ computed for a wave packet
\mbox{$P(v) \cong \exp\left\{ - \frac{[v-v_0(t)]^2}{2 w_0^2} \right\}$}. In contrast to the preceding treatment, the width of $P(v)$ is now fixed according to Eq.~(\ref{w0}). The resulting accelerations, however, remain entirely the same.

In principle, the possible values of the friction coefficient $\gamma$ are only constrained by condition $w_0 \ll  c$, which, according to Eq.~(\ref{w0}), implies $\gamma \ll \langle D_{\text{v}}\rangle / c^2$. 
The friction itself can be related to the fundamental Planck-scale physics, or it can be due to non-universal interactions between the environment of the object. One estimate for the former case is $\gamma \sim 1/\tau_\text{A}$, where $\tau_\text{A}$ is given by Eq.~(\ref{tau1}).

\section{Discussion} \label{discussion}

1) The theoretical framework of our treatment is based on the postulate that the relative velocity of two macroscopic objects acquires definite value only after the two objects measure it, i.e. only after the variable representing the relative velocity becomes entangled with the internal degrees of freedom of the two objects. The same presumably applies to the relative distance and other observables.

2) In this work, we describe the mutual measurements of only one physical observable, namely the relative velocity, governed by one particular mechanism, namely, the modulations of Doppler shifts originating from the fluctuations on the Planck spatial scale. Doppler shift is not the only possible way of the entanglement between the relative velocities of two macroscopic objects. The entanglement can, in principle, also proceed through other mechanisms such as the one described in Appendix~\ref{sec_spreading}, namely, the scattering of photons dependent on the CM position of a macroscopic object, cf. Ref.~\cite{arndt}. Likewise, even if the Doppler shifts are involved, they do not necessarily originate from the Planck-scale fluctuations. Other possible origins of the statistical velocity fluctuations and of the mutual measurements of the distance or other variables require further investigations. The resulting corrections can potentially lead to modifications of the Newton's gravitation~\cite{Milgrom,smolin}.

3) The assumption that the characteristic time $\tau_\text{A}$ given by Eq.~(\ref{tau1}) originates from the fluctuations on a fixed spatial scale $l_0$ implies diffusive random motion in real space. If the parameters of this motion are constrained by the Heisenberg uncertainty, then the relevant diffusion coefficient can be estimated as $D_\text{x} \sim \hbar/M$, where $M$ is either $M_\text{A}$ or $M_\text{B}$, cf. Ref.~\cite{Badiali}.

4) In this work, we did not attempt a relativisticaly invariant treatment. The applicability of our assumptions is, in particular, limited by condition $v \sim \frac{\hbar}{M l_0} \ll c$, which implies that the masses involved must be much larger than the Planck mass $m_P \equiv \sqrt{\frac{\hbar c}{G}} =\frac{\hbar}{cl_0}\approx 2 \ 10^{-5}\,$g.

A related assumption is that the proposed treatment applies only to systems having a very large number of internal degrees of freedom. Therefore, it excludes microscopic particles such as neutrons. However, given that the gravitational field of the Earth does affect quantum phases in neutron interferometry experiments~\cite{Overhauser-74,Colella-75}, the question naturally arises about the compatibility of our proposal with these experiments. Here we observe that the macroscopic apparatus, which is used to prepare the initial state of a neutron and then to measure it, experiences the Earth's gravitational field. Such a setting may be similar to a Gedanken experiment on neutron interference in a uniformly accelerated laboratory, where a neutron uncoupled from the apparatus and left to itself  evolves as a free particle in an inertial reference frame; however, since the apparatus is accelerating, it would detect all quantum effects associated with the presence of the fictitious force in a non-inertial reference frame. If the Earth's gravity is to be described as a fictitious force in an accelerated reference frame, then the counterpart of the inertial reference frame   would  be the one moving with a freely falling body. The internal consistency of the above picture requires further investigation.

5) When defining the mutual measurement mechanism in Section~\ref{single} through the exchange of photons, we neglected the recoil of object A due the finite momentum of a photon $p_{\text{ph}} =\hbar \omega /c$. This assumption is valid as long as $p_{\text{ph}} \ll \Delta p_\text{A}$ , where $\Delta p_\text{A} = M_\text{A} \Delta v_\text{A}$ is the width of object A's momentum wave packet. We can now check this assumption for our gravitation scenario by considering object A with density $\rho_A$ and the characteristic size $R_A$ observed by object B located distance $R_A$ away. In this case, Eqs.~(\ref{resolve2},\ref{tau1}) imply that 
$\frac{p_{\text{ph}}}{\Delta p_\text{A}} \sim \frac{\hbar^2 \omega}{\rho_\text{A}^2 R_\text{A}^5 l_0^2 c^3} $. For one gram of water and an infrared photon corresponding to temperature 300~K, this ratio is $10^{-6}$, and it becomes smaller for larger and/or denser objects.

6) The mechanism behind gravitational attraction considered in this work relies on the exchange of photons between source A and observer B. For the consistency of this mechanism, the observer should detect at least one photon emitted by the source during time interval $\tau_\text{A}$. Let us assume that the source is a black body at temperature $T$ with mass $M_\text{A}$ and surface area $A_\text{A}$, and that the observer detects all photons within solid angle $\Omega$. In such a case, the number of detected photons  per time $\tau_\text{A}$ is
$n_{\text{ph}} \sim \frac{\sigma_{\text{SB}} l_0^2   }{k_B \hbar } T^3 A_\text{A} M_\text{A} \Omega$, where $\sigma_{\text{SB}}$ is the Stefan-Boltzmann constant. 
Such a constraint poses no problem for objects like Earth observed across the solar system. However, the implications of this constraint for smaller objects or very large distances require further analysis. 

We note in this regard that, in order for the observer to become entangled with the source, the observer does not need to receive a photon directly from the source. Instead, the source can be first entangled with a massive nearby object, which would then retranslate the information about the source velocity, amplifying it through a much larger number of emitted photons. We further note that scattering of externally arriving photons by the source is another valid mechanism for the entanglement generation between the source and the observer.

7) Velocity fluctuations with characteristic time $\tau_A$ given by Eq.~(\ref{tau1}) imply random trembling motion for the CM coordinate. Let us consider such a trembling motion for a very large object of the size of Earth in free space. A small-mass observer gravitationally bound to this object would then become entangled with the CM trembling motion. As a result, the observer finds itself in a trembling non-inertial reference frame and starts  perceiving distant vacuum as performing a trembling motion with the same characteristic time. It is an interesting speculation that the consistency of such a treatment would require the observer to detect radiation with characteristic frequency $\hbar/\tau_A$. The temperature corresponding to such a radiation would be 
\begin{equation} \label{T0}
T_A \cong \frac{ \hbar c^3}{k_B G M_A}.
\end{equation}
The right-hand-side above is by factor $8 \pi$ larger than the value of the Hawking radiation temperature for a black hole of mass $M_A$~\cite{Hawking1975}. However, in our analysis, we did not fix the prefactors. 

For the mass of Earth, the estimate~\eqref{T0} gives \mbox{$T_A \sim 0.5\,$K.} Remarkably, this value is only by a factor of 5 smaller than the temperature of the cosmic microwave background (CMB) radiation $T_{\text{CMB}} = 2.72\, $K~\cite{planck2015}. It is, therefore, an interesting possibility that the fundamental random motion of the CM of the Earth arising from the assumptions of the present article is characterized by the temperature $2.72\,$K. An obvious way to falsify the Earth's CM fluctuations as the origin of CMB is to observe CMB with temperature $2.72\,$K in a location in space where the gravitational field of Earth is negligible. However, to the best of our knowledge, no such measurements were made by now. Even in the Planck satellite experiment~\cite{planck2015}, the apparatus was located at the Earth/Sun $L_2$ point, where Earth still makes significant contribution to the local gravitational field. We further note that, if $T_{\text{CMB}}$ were to be associated with the fluctuations of the Earth's CM, then fluctuations of the Sun's CM would correspond to temperature $T_{\text{CMB}} \frac{M_E}{M_S} \approx 8.2\,\mu$K, where $M_E$ and $M_S$ are the masses of Earth and Sun respectively --- giving the black-body radiation maximum at about $0.5$\,MHz. We are not aware of a measured cosmic background peak at that frequency, but interesting excess radiation measurements were reported around frequency 3\,MHz~ \cite{yates,novaco,cane_rad,arcade_excess_1,arcade_excess_2}.

As mass $M_A$ in Eq.~\eqref{T0} decreases, the temperature $T_A$ in Eq.~\eqref{T0} increases, so that for objects of the size of mass 1\,kg, $T_A \approx 3\ 10^{24}$\,K. Such an unrealistically high temperature does not necessarily have observable implications for atoms belonging to the 1-kg object. Firstly, it is just the property of only three degrees of freedom (CM coordinates)  weakly coupled to more than $10^{24}$ remaining degrees of freedom. Secondly, in the Earth setting, each atom belonging to this 1-kg object feels the strong gravitational field of the Earth, while the gravitational field due to the 1-kg object itself is a rather small correction.
Absent a complete theory,  one may speculate that the intensity of the above-described radiation is dominated by  the Earth mass, interaction with which may suppress or distort the fluctuations of the 1-kg object.

8) Our treatment can be compared with the quantum collapse-based gravity models of Diosi-Penrose type~\cite{Diosi-90,Penrose1996,Donadi-20}--- in particular, those considered in Refs.~\cite{Kafri_2014} and~\cite{tilloy}. 
The model described in Ref.~\cite{Kafri_2014} is based on the assumption that gravity introduces unconventional communication  channels between objects on the top of those describable within the Standard Model. These channels are responsible for quantum decoherence and measurements. Hence the discussion of the models often involves the notions of   gravitation-induced decoherence and gravitation-induced collapse. In contrast, the present work describes  gravitation as emerging from the communication based on the Standard Model mechanisms, e.g. through photons. No unconventional decoherence is anticipated. The only ingredient beyond the Standard model is the postulated fluctuations on the Planck spatial scale.
In Refs.~\cite{Kafri_2014} and~\cite{tilloy},  the  models are based on quantum position measurements, while our analysis is based on the velocity measurements. In these models, the process of measurement and feedback was postulated in the framework of toy models without specifying the underlying mechanisms.  In the present work, the measurements are  specifically based on registering Doppler shifts, and simultaneously, the Doppler effect is also the part of a concrete feedback mechanism, where the rate of the mutual velocity measurements depends on the relative velocities of the two objects. On the other hand,  we have not yet proposed a concrete model of the Doppler-shift-based measurements and feedback.

\section{Conclusions} \label{sec_conc}
In the present article, we defined the concept of mutual quantum measurements of two macroscopic objects and showed that mutual measurements of velocity can lead to an emergent force acting on the objects. We further showed that, under a set of assumptions involving the measurements of Doppler shifts originated from fluctuations on the Planck scale, the measurement-induced force can reproduce Newton's law of gravitation.  The resulting framework guarantees the equivalence between the inertial and the gravitational masses. Under a broader set of assumptions, emergent forces and other dynamical effects of quantum measurements can lead to corrections to Newton's law of gravitation.

\acknowledgments
The authors are grateful to Lajos Diosi, Oleg Lychkovskiy and Alexander Rozhkov for valuable feedback on the manuscript. This work was supported by a grant of the Russian Science Foundation (Project No. 17-12-01587).

\appendix

\section{Relevant foundational issues of quantum mechanics}
\label{issues}

In this Appendix, we summarize our views on two foundational issues of quantum mechanics related to the operational definition of mutual quantum measurement given in Section~\ref{mutual}. 

\subsubsection{Use of a single realization of the quantum evolution}

Let us recall that the statistical interpretation of quantum mechanics stipulates that a quantum description, such as the one starting from the wave function~(\ref{Psi1}) and ending with the density matrix~(\ref{rhof}) describes the outcome of the ensemble of measurements characterizing many repeated experiments starting from the same initial quantum state~(\ref{Psi1}). The diagonal elements of the density matrix~(\ref{rhof}) then give the probabilities of the individual states $ |f_1S_1\rangle$, and $ |f_2 S_2\rangle$ to be observed as a result of a measurement. For small systems, such an ensemble of measurements can be, indeed, realized experimentally or naturally. However, for large systems, the size of the ensemble required to test the quantum mechanical predictions is exponentially large in terms of the number of particles in the system, which, presumably, makes such an ensemble not realizable either experimentally or, more importantly, naturally.

The above considerations imply that the exact quantum-mechanical state of a typical macroscopic object at any moment of time is not fully repeatable in the future.  Therefore, we think it is reasonable to extend the quantum-mechanical formalism to a single realization of the time evolution of macroscopic objects~\cite{kent1}. Here, one is helped by the results on quantum typicality in many-body systems~\cite{gemmer,goldstein,bartsch,popescu},  which suggest that, in numerous settings, averages obtained from a single realization of a many-body wave function randomly chosen from the relevant statistical ensemble are exponentially close to the averages for the entire ensemble.

\subsubsection{Defining macroscopically distinct quantum states}

The two quantum states denoted in Section~\ref{single} as $|S_1 \rangle$ and $|S_2 \rangle$ and describing the internal degrees of freedom of object B are assumed to lose their quantum coherence very quickly. We expect that these two states are macroscopically distinct either from the very beginning - if the absorption of the large number of photons was involved in their creation, or - if only one photon is involved - they quickly become macroscopically distinct via the process of the propagation of perturbations in an ergodic quantum system~\cite{absence_exp,tarek_large,Maldacena2016,susskind}.

It is well known that classifying two quantum states as macroscopically distinct as opposed to being non-macroscopically distinct is a notoriously difficult subject~\cite{leggett,Korsbakken_2009,leggett2,gisin}. One can argue here that any pair of two orthogonal quantum states for an isolated system is as good as any other orthogonal pair. It is probably impossible to define a distance between two orthogonal quantum states, which would be invariant under the Hilbert-space rotations and, at the same time, distinguish macroscopically different from non-macroscopically different orthogonal states.  

In the present article, we, therefore, adopt the following practical definition: we call two quantum states of a given system ``macroscopically distinct'', if, according to the standard quantum mechanics, the quantum coherence time of the superposition of these two states in a given environment is smaller than a practically unmeasurable value, e.g., $10^{-43}$s (the Planck time).

\section{Alternative mechanism for measuring the center-of-mass momentum} \label{sec_spreading}

The mutual velocity measurements described in Section~\ref{mutual} can occur not only via the detection of Doppler shifts. In this Appendix, we outline one alternative.

\begin{figure}[t]
 \centering
 \includegraphics[width=0.9\columnwidth]{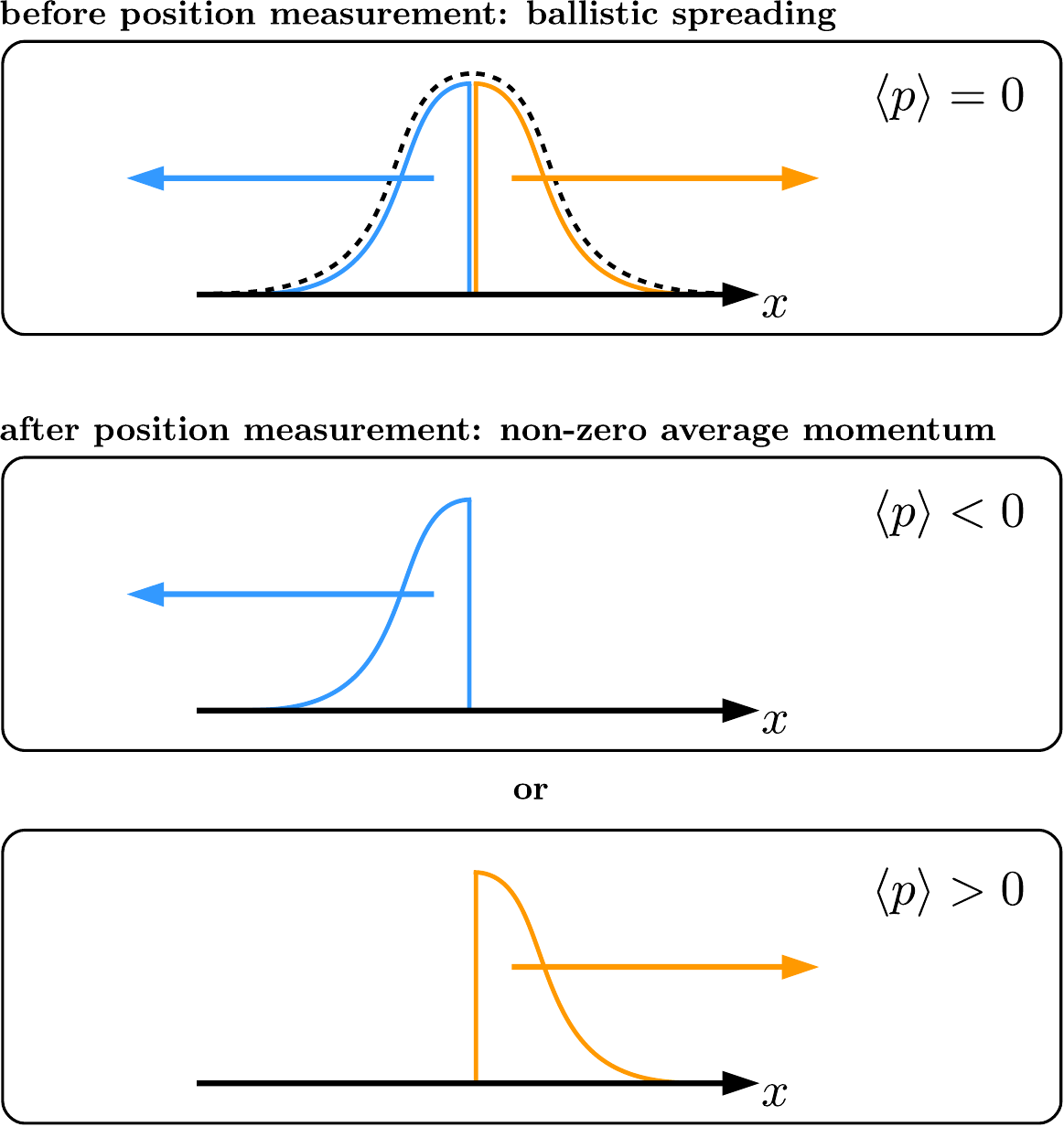}
 \caption{Illustration of the position-momentum correlations for spreading wave packets after position measurements.}
 \label{fig_pm}
\end{figure}

Let us consider the quantum-mechanical wave packet describing the CM position of an object of mass $M$. According to the Schr{\"o}dinger equation for a free massive particle, this spatial wave packet spreads ballistically between two subsequent measurements. For example, a wave packet, which initially has the Gaussian shape $\psi(x)\cong\exp[-x^2/(2\sigma_0^2)]$ with width $\sigma_0$, broadens in time such that its time-dependent width reads~\cite{greiner_qm}
\begin{equation} \label{eqn_br}
 \sigma(t)=\sqrt{\sigma_0^2+(v_\text{sp}t)^2},
\end{equation}
where $v_\text{sp}\equiv\hbar/(M\sigma_0)$ is the spreading velocity. This spreading leads to a correlation between the CM position and CM momentum, as illustrated in Fig.~\ref{fig_pm}. Indeed, let us assume that the initial average momentum of the wave packet $\langle p \rangle$ is zero, as in the above example. Therefore, the average CM position under Schr{\"o}dinger time evolution also remains zero. If, at a later time, the CM position is measured and found to be on the right of the initial average position, then this measurement picks up the momentum component directed to the right, and {\it vice versa}. Such a measurement, can be realized when a photon wave packet of a size smaller than $\sigma(t)$ is scattered from the massive object and then detected.

\section{Need for new physics on the Planck scale?}
\label{Planck}

In Section~\ref{PlanckLength}, we estimated the fluctuation time on the Planck spatial scale with the help of Eq.~(\ref{tau1}) relying on non-relativistic quantum mechanics. The natural question then arises about the meaningfulness of such an extrapolation to the scale, where quantum mechanics has not been tested. Here we rely on the consideration that, even though we deal with the CM wave packets of very small widths, the matter is not really concentrated within these widths. The CM wave packet of a macroscopic object represents the CM position of many constituting particles described by regular quantum mechanics. It may even happen that no physical matter is located within the width of the CM wave packet as, for example, would be the case for a hollow sphere. At the same time, the relevant velocities are very small. This consideration not only supports the extension of non-relativistic quantum mechanics to the Planck scale but also hints that the use of quasi-classical approximation for the resulting dynamics, which would involve the notion of classical chaos~\cite{milburn}, Lyapunov exponents, etc., may be reasonable.

Another question is whether the fluctuations on the Planck scale, which were assumed to justify Eq.~(\ref{tau1}), require a fundamentally new force field. Such a need is not clear from the perspective of the present work. The deviation from the coherent Schr\"{o}dinger dynamics on the Planck scale may originate from the cumulative influence of the rest of the Universe on a given system via routine mechanisms known for the larger spatial scales, such as classical forces or quantum measurements. 

If the new force exists, it might be vanishingly small, which would make the situation somewhat similar to the one associated with the role of the rounding errors in the simulations of classical many-body chaotic systems on a computer~\cite{wijn_lyapunov,wijn_lyapunov2}. These errors originate from the finite precision of internal computer operations. Since the chaotic dynamics exponentially amplifies the effects of vanishingly small changes of the simulated coordinates, it also amplifies the rounding errors. Yet, for high-dimensional phase spaces, the statistical behavior of the error-modified chaotic trajectories on practically computable timescales remains almost indistingishable from that of the error-unmodified trajectories, because the rounding errors introduce only a tiny change of the values of the Lyapunov exponents. One becomes aware of the computer rounding errors only after attempting to simulate perfect time reversal of the dynamics~\cite{absence_exp,tarek_large,Tarkhov_2018}. Likewise, the additional physics on the Planck and sub-Planck scales may add negligible correction to the processes controlled by the many-body entanglement and decoherence, while making qualitative difference only in the suppression of the perfect time reversal. 

At the same time, it cannot be excluded that the velocity fluctuations on the Planck scale postulated in this work originate from the new physics, such as, for example, the inhomogeneous character of the physical space on that scale.

\section{Mutual measurements of more than two objects: an example}
\label{morethantwo}

The preceding section highlights the following general question: Can the scheme presented in Sections~\ref{multiple} and \ref{assumptions} for the mutual measurements of two objects be consistently extended to more than two objects? While the detailed investigation of this question is beyond the scope of the present paper, here we discuss one relevant example.

According to Eqs.~(\ref{resolve2}) and (\ref{tau1}), when object $B$ measures object $A$,  the variance of the velocity fluctuations for this measurement is
\begin{equation}
    \langle \Delta v_\text{A}^2 \rangle = \alpha^2 M_\text{A}^2 ,
\label{dvA2}
\end{equation}
where $\alpha \cong \frac{l_0^2 c^2}{\hbar r}$.

Let us now consider what happens, if object $A$ is divided into two parts $A_1$ and $A_2$ with equal masses $M_\text{A}/2$, and then each of the two parts is measured separately by object $B$. In that case, 
$\langle \Delta v_{A_1}^2 \rangle = \langle\Delta v_{A_2}^2 \rangle  = \alpha^2 M_\text{A}/4$.
On the other hand, since $v_\text{A} = (v_{A_1} + v_{A_2})/2$, then also $\Delta v_\text{A} = (\Delta v_{A_1} + \Delta v_{A_2})/2$. If $\Delta v_{A_1}$ and $\Delta v_{A_2}$ are statistically independent, then $\langle \Delta v_A^2 \rangle = (\langle \Delta v_{A_1}^2 \rangle + \langle \Delta v_{A_2}^2 \rangle)/4 = \alpha^2 M_\text{A}/8$, which is a factor of 8 smaller than the result in Eq.~(\ref{dvA2}). How can this contradiction be resolved?

The answer is that objects $A_1$ and $A_2$ are measuring each other , and this modifies the information received by object $B$. 

Let us assume that, during each interval $\tau_\text{A}/2$, object $B$ registers the change of the velocity of $A_1$ from its own direct measurements, $\Delta v_{A_1}$ of $A_1$,   and simultaneously it receives an independent information about the  relative velocity fluctuation $\delta_2v_{A_1}$ of $A_1$ as measured by the internal degrees of freedom of $A_2$. Now, taking into account that object $B$ also registers the velocity change $\Delta v_{A_2}$ of $A_2$  directly, it updates the velocity of $A_1$ as
\begin{equation}
v_{A_1} \to v_{A_1} + \Delta v_{A_1} + \Delta v_{A_2} + \delta_2v_{A_1}.
\label{vAp}
\end{equation}
Likewise, it updates the velocity of $A_2$ as
\begin{equation}
v_{A_2} \to v_{A_2} + \Delta v_{A_2} + \Delta v_{A_1} + \delta_1v_{A_2}.
\label{vApp}
\end{equation}
As a result the recorded correction to the center of mass velocity of $A_1$ and $A_2$ becomes $\Delta v_\text{A} = (2 \Delta v_{A_1} + 2 \Delta v_{A_2} + \delta_2v_{A_1} + \delta_1v_{A_2})/2$.
The total momentum conservation in the mutual measurements of $A_1$ and $A_2$ implies that $\delta_2v_{A_1} + \delta_1v_{A_2} =0 $. Thus we obtain $\langle \Delta v_\text{A}^2 \rangle =  \alpha^2 M_\text{A}/2$. This is still a factor of 2 smaller than the right-hand side in Eq.~(\ref{dvA2}). Here, we recall however, that, according to our assumption the variance in Eq.~(\ref{dvA2}) is accumulated over time $\tau_\text{A}$, while the factor-of-two smaller value obtained in this paragraph is accumulated over time  $\tau_\text{A}/2$, which, therefore, implies the consistency of the two expressions.


%

\end{document}